# Can Bayes Factors "Prove" the Null Hypothesis?

Michael Smithson, The Australian National University

Email: Michael.Smithson@anu.edu.au

Abstract

It is possible to obtain a large Bayes Factor (BF) favoring the null hypothesis when both the null and alternative hypotheses have low likelihoods, and there are other hypotheses being ignored that are much more strongly supported by the data. As sample sizes become large it becomes increasingly probable that a strong BF favouring a point null against a conventional Bayesian vague alternative co-occurs with a BF favouring various specific alternatives <u>against the null</u>. For any BF threshold $q$ and sample mean $\bar{X}$, there is a value $n$ such that sample sizes larger than $n$ guarantee that although the BF comparing $H_0$ against a conventional (vague) alternative exceeds $q$, nevertheless for some range of hypothetical $\mu$, a BF comparing $H_0$ against $\mu$ in that range falls below $1/q$. This paper discusses the conditions under which this conundrum occurs and investigates methods for resolving it.

One attraction of Bayesian statistical methods for researchers is the belief that, unlike the *p*-value in frequentist null-hypothesis tests, a large Bayes factor (BF) in favour of the null hypothesis and against its alternative is strong evidence for the null hypothesis (Gallistel, 2009; Masson, 2011). While well-informed users of Bayesian statistics may be aware that there is more to it than this, it seems that many users may not be. Unfortunately, especially in large-sample research with small effects, this belief can be outright misleading. With increasing sample size, a substantial BF favouring the null hypothesis against a conventional vague alternative will increasingly co-occur with substantial BFs favouring a range of other alternatives <u>against</u> the null hypothesis. Thus, depending on the choice of alternative, Bayes factors may sharply contradict each other regarding the evidentiary status of the null hypothesis. I provide a guide to the conditions under which relying on the Bayes factor to assess the null can mislead researchers and research consumers, and develop practical suggestions for researchers.

Collectively, Bayesians provide diverse advice on how to deal with point null hypotheses for continuous parameters. Some advocate using the Bayes factor, $B_{01}$, comparing a null hypothesis $H_0$ against its complement $H_1$ as evidence for or against the null (Dienes, 2014; Konijn, et al. 2015; Rouder, et al. 2018; Wagenmakers, 2007). The Jeffreys table (Jeffreys, 1961) linking adjectives such as "substantial" and "strong" with ranges of BFs has been popular in this connection. Other Bayesians do not endorse BFs for continuous parameters[8,9,10] (Gelman, ete al. 2004; Lynch, 2007; Liu & Aitkin, 2008), with some recommending using credible intervals instead for making decisions about the null (Kruschke, 2011; Bolstad & Curran, 2016). Still others have pointed out that the Bayes factor and credible interval provide different criteria for assessing the null (Rouder, et al. 2018), along similar lines to the classic Lindley paper highlighting situations where a strong BF favouring the null co-occurs with a frequentist significance test rejecting it (Lindley, 1957). Those in this last camp recommend not using credible intervals to decide whether or not to reject the null.

Setting aside the use of credible intervals, it is easy to forget that the Bayes factor tells us how more or less likely one hypothesis is than another hypothesis, but does not inform us about the absolute likelihoods of either hypothesis.  Thus, it is possible to obtain a large BF when both the null and alternative hypotheses have low likelihoods, and there are other hypotheses being ignored that are much more strongly supported by the data.

We shall see that this is what can happen to a pointwise $H_0$ and vague $H_1$ under certain conditions as sample sizes become large. As sample sizes become large it becomes increasingly probable that a strong BF favouring H$_0$ against a conventional Bayesian vague H$_1$ co-occurs with a BF favouring various specific alternatives <u>against H$_0$</u>. Thus, we may simultaneously find strong support for and against the null from the same data. I will call these co-occurrences "Lindley cases", after the so-called Lindley (1957) "paradox".

Lindley cases are most likely to occur when there are small effects and large samples. For a simple example, consider a two-outcome process producing E or ~E on every trial, and let $\theta$ denote the proportion of trials yielding E. Let us assume that $\theta$ is a binomial variable, setting H$_0$ is $\theta = 1/2$, H$_1$ is $\theta \neq 1/2$, and assigning prior probabilities $\pi(H_0) = \pi(H_1) = 1/2$. Suppose we observe $n$ trials and $k$ occurrences of E. The likelihood of these data given H$_0$ is

$$P(k/n \mid H_0) = \binom{n}{k} 0.5^k 0.5^{n-k}, \tag{1}$$

the likelihood of the data given H$_1$ is

$$P(k/n \mid H_1) = \int_0^1 \binom{n}{k} \theta^k (1-\theta)^{n-k} \, d\theta = 1/(n+1), \tag{2}$$

and the Bayes factor for H$_0$ against H$_1$ is

$$B_{01} = P(k/n \mid H_0) \Big/ P(k/n \mid H_1) = (n+1) \binom{n}{k} 0.5^n. \tag{3}$$

For example, if $n = 985$ and $k = 524$ (so the sample proportion is 0.532) then $B_{01} = 3.344$, which in the Jeffreys table lands in the "substantial" range, 3-10, favouring H$_0$. However, if we examine Bayes factors, $B(\theta, \theta_0, x)$, comparing certain point-valued alternatives against the null in the neighbourhood from 0.5 to 2*(524/985) – 0.5 = 0.564, we observe that these $B(\theta, \theta_0, x) > 1$. Moreover, for hypothetical $0.51042 < \theta < 0.55349$, $B(\theta, \theta_0, x) > 3$, i.e., a Bayes factor greater than 3 against the null. This is a Lindley case. It occurs because both likelihoods for H$_0$ and H$_1$ are relatively small: $P(k/n \mid H_0) = 0.003399$ and $P(k/n \mid H_1) =$ 0.00101, whereas likelihoods in the neighbourhood from 0.5 to 0.564 are higher than either of these.

Introducing another example, we will now focus on scenarios in which we take a sample of size *n* from a normal distribution N ($\mu$, $\sigma^2$) population with known variance $\sigma^2$, testing whether or not the null hypothesis $H_0 : \mu = \theta$ on the mean holds (against the alternative $H_1 : \mu \neq \theta$). Without loss of generality, we shall set $\sigma^2 = 1$ and we also set $\theta = 0$. The Bayesian framework requires specifying what is meant by $H_1$, and for this we will examine more than one version. The first of these is Robert's (2014) Bayes factor for the null against the alternative when $\sigma^2 = 1$:

$$R(\bar{X}, n) = \sqrt{n+1} \exp\left(-\frac{n^2 \bar{X}^2}{2n+2}\right) \tag{4}$$

We can readily confirm the claim by Robert that this equation yields $R(1.96/\sqrt{16,818}, 16,818) \approx 19$ in favour of the null. This reproduces the figures from Lindley's (1957) example, where he shows that a *t*-value of 1.96, sufficient for frequentists to reject the null at $\alpha = .05$, also yields a BF of 19 favouring the null. A bit of algebra reveals that to achieve a desired Robert BF of *q* we require

$$\bar{X} = \frac{\sqrt{(n+1)(\log(n+1) - 2\log(q))}}{n} \tag{5}$$

Two other versions of $H_1$ considered in this paper are from the JZS-scaled and information-scaled Bayes factors generated by the Rouder et al. (2009) t-test. The Robert BF is handy for its tractability, and the Rouder et al. BFs are included here mainly for the purpose of demonstrating the fact that our results generalize to some popular null-hypothesis BFs.

We now move to examining the behaviour of the Bayes factor when the null is compared against a point-valued alternative. A pointwise comparison between two alternatives gives a BF of

$$B(\mu_1, \mu_2, \bar{X}, n) = \exp\left[n\left((\bar{X} - \mu_2)^2 - (\bar{X} - \mu_1)^2\right)/2\right] \tag{6}$$

The log of this is linear in $n$ and the rate of change in $n$ is positive when $|\bar{X} - \mu_2| > |\bar{X} - \mu_1|$ and negative when $|\bar{X} - \mu_2| < |\bar{X} - \mu_1|$. Setting $\mu_1 = 0$, to achieve $\log(B(0, \mu, \bar{X}, n)) = \log(q)$, again straightforward algebra gives two solutions:

$$\mu = \bar{X} \pm \frac{\sqrt{n\bar{X}^2 - 2\log(q)}}{\sqrt{n}} \tag{7}$$

Given $\bar{X} > 0$, say, for any $\mu$ such that $0 < \mu < 2\bar{X}$, $B(0, \mu, \bar{X}, n) > 1$. By contrast, for any $\mu < 0$ or $> 2\bar{X}$, $B(0, \mu, \bar{X}, n) < 1$. Given any criterial threshold, $q$, from equation (7) we can find the sets of $\mu$ satisfying $B(0, \mu, \bar{X}, n) \leq 1/q$ and $B(0, \mu, \bar{X}, n) \geq q$. Moreover, we can use equation (5) to find $\bar{X}$ to input into equation (7) such that $R(\bar{X}, n) \geq q$. We now have the wherewithal to ascertain the conditions under which Lindley cases occur (i.e., combinations of $q$, $n$, and $\bar{X}$ where $B(0, \mu, \bar{X}, n) \geq q$ and $R(\bar{X}, n) \geq q$). We will focus on large samples, because the impact of large samples seems to have been under-explored.

<u>Bayes Factor Example</u>

Suppose $n = 5{,}000$, and our BF threshold is $q = 3$. Then from equation (2) we have $\bar{X} = 0.035557$. Suppose also that our BF threshold for $q$ is 3. Inputting $n = 5000$, $q = 3$ and $\bar{X} = 0.035557$ into equation (7) yields lower and upper bounds $\mu \in [0.0068368, 0.0642769]$. That is, $B(0, \mu, \bar{X}, n) > 3$ against the null when when $0.0068368 < \mu < 0.0642769$. Figure 1 displays a graph of $\log(B(0, \mu, \bar{X}, n))$ as a function of $\mu$, with two horizontal lines demarcating $\log(3)$ and $\log(1/3)$. The vertical lines mark out the two pairs of bounds that have just been derived.

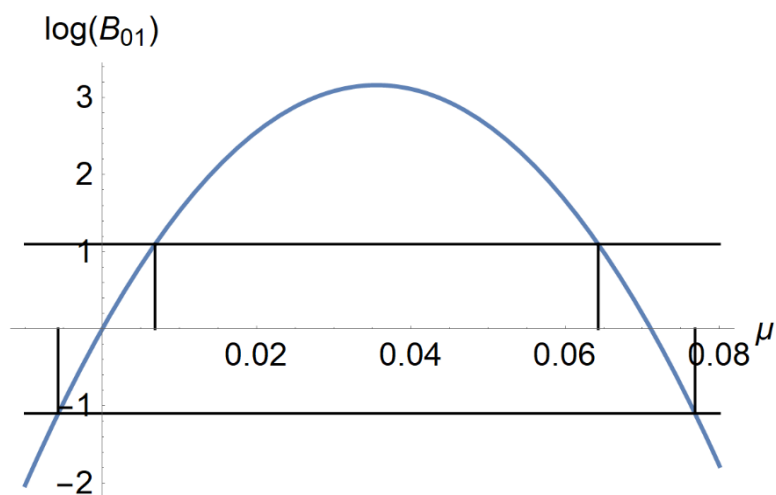

Figure 1. Log of Bayes Factor as a function of $\mu$

In this example, we have seen that under a set of conditions for which the Robert (2014) BF $R(\bar{X}, n) = 3$ in favour of the null hypothesis, there is a neighbourhood around $\bar{X}$ for which a comparison between the null and every alternative value of $\mu$ in this neighbourhood yields a BF greater than 3 against the null. How does the Rouder et al. (2009) Bayesian t-test fare here? The requirement that $\bar{X} = 0.035557$ to attain a BF threshold of 3 is equivalent to requiring a *t*-value of approximately $t = 0.035557\sqrt{5000} = 2.514$. Pursuing this with the JZS and scaled-information BFs from the Rouder test, it turns out that the *t* values required to obtain BFs $\geq 3$ are 2.467 for the JZS and 2.373 for the scaled-information tests. Thus, the required sample means are $\bar{X} = 2.467/\sqrt{5000} = 0.03489$ for the JZS and $\bar{X} = 2.373/\sqrt{5000} = 0.03356$ for the scaled-information tests. These yield similar results to the preceding outcome using the Robert BF, i.e., a neighbourhood around $\bar{X}$ for which a comparison between the null and every alternative value of $\mu$ yields a Bayes factor greater than 3 against the null. These neighbourhoods are [0.00700, 0.06278] for the JSZ and [0.00735, 0.05977] for the information-scaled, both of which are fairly similar to the Robert-BF based interval derived above.

How Do Lindley Cases Arise?

A measure of evidence related to Bayes factors is the "evidence ratio" (Morey et al., 2016; Wagenmakers, et al. 2019), which can be written in two ways:

$$ER(\theta, x) = \frac{p(\theta|x)}{p(\theta)} = \frac{p(x|\theta)}{p(x)}. \tag{8}$$

The evidence-ratio compares the conditional probability of $\theta$ in the posterior distribution with its probability in the prior. If the same prior is used as the vague "alternative" against which to compute a Bayes factor for a point null, then the evidence-ratio at that point is the Bayes factor. Evidence-ratios greater than 1 favor the hypothetical $\theta$, and taking this idea one step farther yields the concept of a "support interval" (Wagenmakers, et al. 2019), the values of $\theta$ for which $ER(\theta, x)$ exceeds some criterial value $q$. We may also take the analogous step of declaring a "rejection region" as the collection of values of $\theta$ for which $ER(\theta, x)$ falls below the criterial value $1/q$. Thus, support and rejection regions for $\theta$ are tantamount to regions displaying the point "nulls" with Bayes factors large or small enough to decide to retain or reject the "nulls".

There also is a straightforward connection between Bayes factors comparing a fixed point null with point-valued alternatives and the evidence-ratios for these alternatives. In equation (8), the denominator of the right-most ratio is a marginal likelihood and therefore a constant that does not involve $\theta$. A BF that compares a point alternative $\theta$ against a point-null $\theta_0$ may be written as

$$B(\theta, \theta_0, x) = \frac{p(x|\theta)}{p(x|\theta_0)}, \tag{9}$$

which therefore is proportional to the evidence-ratio for $\theta$. Their ratio is

$$\frac{ER(\theta, x)}{B(\theta, \theta_0, x)} = \frac{p(x|\theta_0)}{p(x)} = ER(\theta_0, x), \tag{10}$$

the evidence-ratio for $\theta_0$. As mentioned above, if the prior involved in $ER(\theta, x)$ is the same as the vague alternative employed in a conventional point-null Bayes factor $B_{01}$, then $ER(\theta_0, x) = B_{01}$, regardless of the fact that $B(\theta_0, \theta_0, x) = 1$. Equation (10) implies that setting a criterial threshold for $ER(\theta_0, x)$ to determine a support region also sets a criterial threshold for $B(\theta, \theta_0, x)$, although the latter threshold is not known until the data have been collected. The converse also holds.

A source of utility in this connection is that both evidence-ratios and Bayes factors have clear and complementary betting interpretations. The evidence-ratio gives us the betting-odds $ER(\theta, x)$ on $\theta$ relative to the posterior distribution against the prior. The Bayes factor $B(\theta, \theta_0, x)$, on the other hand, gives the betting-odds on $\theta$ against $\theta_0$. Moreover, the relationship between $ER(\theta, x)$ and $B(\theta, \theta_0, x)$ provides an explanation for the Lindley-case syndrome.

Returning to the binomial example earlier in this paper, recall that H0 is $\theta_0 = 1/2$ and data are $n = 985$ and $k = 524$ so the sample proportion is 0.532. Earlier we ascertained that $B_{01} = 3.344$. If we use the uniform prior then it also is the case that $ER(\theta_0, x) = 3.344$, and this is the evidence-ratio-Bayes-factor ratio in equation (10). If we set a threshold-value for evidence-ratios of $ER(\theta, x) > 3$ then the support region is [0.499, 0.565]. This interval includes $\theta_0 = 1/2$, and so we find that the data support the null. Now, recall that for hypothetical $0.51042 < \theta < 0.55349$, $B(\theta, \theta_0, x) > 3$ <u>against</u> the null. Thus, for $\theta$ in this interval the evidence ratios $R(\theta, x) > 3*3.344 = 10.032$. Therefore, the Lindley-case syndrome arises because the $\theta$ in [0.51042, 0.55349] are much more strongly supported by the data than the null, $\theta_0 = 1/2$.

In our Gaussian example with $n = 5000$ and sample mean 0.035557, we found a $B(0, \mu, \bar{X}, n) \geq 3$ <u>against</u> the null for comparisons with any specific $\mu$ in [0.0068368, 0.0642769]. In this example, $ER(\mu_0, x) = 3$ when $\mu_0 = 0$. The evidence-ratios for these interval limits therefore are 9. Again, the Lindley-case syndrome arises because the $\mu$ in [0.0068368, 0.0642769] are more strongly supported by the data than $\mu_0 = 0$.

We now further examine the conditions under which Lindley cases occur. To begin, let us consider the range of $\mu$ values for which $B(0, \mu, \bar{X}, n) > q$ against the null at the same time as the Robert (2014) Bayes factor $R(\bar{X}, n) > q$ as a function of sample size. For any given $\bar{X}$ as $n$ increases the $B(\bar{X}, 0, \mu, n) > q$ range also will expand and $R(\bar{X}, n)$ also will increase. We can see this by differentiating the second term on the right-hand side of equation (7), which gives

$$\partial \frac{\sqrt{n\bar{X}^2 - 2\log(q)}}{\sqrt{n}} \Big/ \partial n = \frac{\log(q)}{n^{3/2}\sqrt{n\bar{X}^2 - 2\log(q)}} \qquad (11)$$

This is negative for log(q) < 0 and positive for log(q) > 0 when $n\bar{X}^2 > 2\log(q)$. Thus, when $B(0, \mu, \bar{X}, n) > q$ the limits grow further apart as $n$ increases. Figure 2 illustrates this by comparing the $\log(B(0, \mu, \bar{X}, n))$ as a function of $\mu$ for $n = 5,000$ and 10,000. The shallower curve is for the smaller sample size. As in Figure 1, the horizontal lines demarcate log(3) and log(1/3).

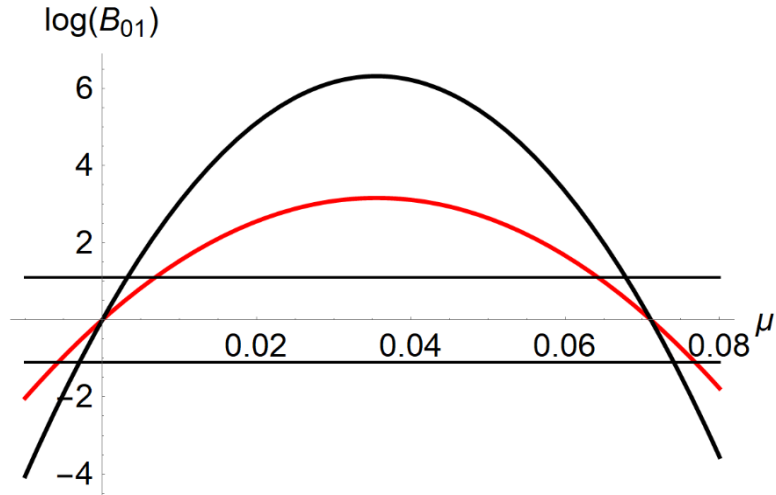

Figure 2. Log of Bayes factor as a function of for $n = 5,000$ and $10,000$

Given $n$ and $q$, if $R(\bar{X},n) = q$ for some $\bar{X}$ then for $\bar{X}$ closer to 0 $R(\bar{X},n) > q$. However, $B(0,\mu,\bar{X},n) > q$ holds only for values of $\mu$ in the neighbourhood around $\bar{X}$ as explained earlier. For every $\bar{X}$ in the appropriate range, there is a minimum $n$ required to satisfy $B(0,\mu,\bar{X},n) > q$. So a reasonable question is how the minimum $n$ covaries with $\bar{X}$ throughout this range. For any benchmark sample size, $n_b$, we can establish the subrange of $\bar{X}$ in which the minimum required $n$ is exceeded by the benchmark $n_b$. That will then give us the range of $R(\bar{X},n)$ for which, given $n$ and $q$, it also is the case that $B(0,\mu,\bar{X},n) > q$ for at least some values of $\mu$ around $\bar{X}$.

In the example, $n = 5000$ and $q = 3$, so inverting equation (5) gives $\bar{X} = \sqrt{2\log(q)/n} = 0.0209629$. At that value of $\bar{X}$, $R(\bar{X},n) = 23.5778$, so this is the upper bound on $B_{R01}$ when $n = 5000$ and $q = 3$ for which there is a neighbourhood of $\mu$ in which $B(0,\mu,\bar{X},n) > 3$. The resulting subrange of $\bar{X}$ is [0.0209629, 0.035557], and these are the sample means for which $3 \leq R(\bar{X},n) \leq 23.5778$. Thus, for a sample of 5000 obtaining a Robert BF anywhere in the range $3 \leq R(\bar{X}, n) \leq 23.5778$ also implies that $B(0, \mu, \bar{X}, n) > 3$ for at least some values of $\mu$ around $\bar{X}$, thereby constituting a Lindley case.

Note that we obtain similar subranges of $\bar{X}$ for the Rouder et al. t-tests, with the same lower limit as the one just derived. Recall that the upper limits of $\bar{X}$ for which the JSZ and scaled-information BFs exceed 3 are 0.03489 and 0.03356 respectively. At the lower limit of $\bar{X} = 0.0209629$, the JSZ BF = 20.9126 and scaled-information BF = 16.6752. Thus, for instance, given a sample of 5000 and JSZ BF between 3 and 20.9126, we know that it will be a Lindley case.

Our next concerns are threefold. First, how does this Lindley case region for $R(\bar{X}, n)$ or the Rouder et al. t-test BFs behave as $n$ increases? Second, for any particular $n$, how likely is it that even when $H_0$ is <u>true</u>, nonetheless Lindley cases will occur? Third, how does that probability covary with $n$ or $q$?

Starting with the first concern, equation (11) tells us that as $n$ increases the threshold $\bar{X}$ will decrease. Substituting the inverted version of equation (11) into equation (4) gives this version of $R(\bar{X}, n)$:

$$R(\bar{X}, n) = \sqrt{n+1} \exp\left(-\frac{n 2 \log(q)}{2n+2}\right) \tag{12}$$

Differentiating it with respect to $n$ yields

$$\frac{\partial R(\bar{X}, n)}{\partial n} = \frac{q^{\frac{1}{n+1}-1}(n - 2\log(q) + 1)}{2(n+1)^{3/2}} \tag{13}$$

which is positive for $n > 2\log(q) - 1$. Therefore, although the threshold $\bar{X}$ declines with increasing $n$, the upper bound on $R(\bar{X}, n)$ nevertheless increases.

To address the second concern, let us return to our example. We have $n = 5000$ and $q = 3$, and inverting equation (5) gives $\bar{X} = \sqrt{2\log(q)/n} = 0.0209629$. As observed earlier, the upper value of the appropriate range for $R(\bar{X}, n)$ is $\bar{X} = 0.035557$. Likewise there is a corresponding interval below 0, [-0.035557, -0.0209629]. The standard error of the mean is $1/\sqrt{n}$, so it turns out that the areas under the normal curve for these two intervals sum to approximately 0.1263. Thus, the probability of simultaneously confirming and disconfirming $H_0$ (i.e., observing a Lindley case) <u>when $H_0$ is true</u> is 0.1263. The same argument applies to the Rouder et al. t-tests. The areas of their intervals under the normal curve turn out to be 0.1246 for the JZS and 0.1206 for the scaled-information tests. These are not negligible probabilities.

Regarding the third concern, the probability of observing a Lindley case when $H_0$ is true increases with *n*. For $R(\bar{X}, n)$, at *n* = 10,000 the probability is 0.130 and at *n* = 20,000 the probability is 0.133, and it is easy to show that it asymptotes at about 0.138. In general, the asymptote is $2 - \text{erfc}\left(-\sqrt{\log(q)}\right)$. On the other hand, we can see that it decreases as *q* increases (e.g., when *q* = 10 the asymptote is 0.032). Nevertheless, these findings suggest that for values of *q* regarded by many researchers as "substantial", the probability of Lindley cases is non-negligible even when $H_0$ is true, and moreover it increases with sample size.

Therefore, point nulls are not "provable" by Bayes factors. Instead, if we want to find the "best bet" on $\theta$ then examining $B(\theta, \theta_0, x)$ (or at least the full range of $ER(\theta, x)$) makes more sense than restricting attention to $ER(\theta_0, x)$ or the corresponding BF.

Quandary at the Boundary

Unfortunately, it is not difficult to show that with large samples support-rejection regions can become problematic in an additional way to Lindley cases. Given a big enough sample, a value arbitrarily close to the point null may be "rejected" by support-interval logic while the point-null will be "supported" at the same criterial ratio. For instance, given a sample size of 6,193 from a Bernoulli random variable, a sample proportion of 0.5156 "rejects" $\theta = 0.495$ and "supports" $\theta = 0.5$ at a criterial evidence-ratio of $q = 3$.

Returning to our scenario of a normal random variable, for simplicity let us assume a prior distribution for the mean of $N(0,1)$ with a weight that counts it as 1 observation. Then the log of the evidence-ratio for an hypothetical $\mu$ is

$$L(\bar{X}, \mu, n) = \left(\bar{X}^2 - n(\bar{X} - \mu)^2 + \log(n)\right)/2. \qquad (15)$$

Given $\mu = 0$, $n$, and a target log-ratio $\log(q)$, the sample mean required to achieve this target is

$$\bar{X} = \pm\sqrt{(\log(n) - 2\log(q))/(n-1)}. \qquad (16)$$

Note that this is very close to the sample mean identified in equation (5) for the target Robert BF (we are using a slightly different prior from his). For example, given $\mu = 0$, $n = 5000$, and target $q = 3$, the required sample mean is $\pm 0.035556$.

Symmetry permits us to deal with just the positive mean, so we do so from here on. Given $n$, target $q$, and the positive version of the sample mean defined in equation (16), we may solve equation (15) to find the hypothetical means required for the target to be $1/q$, which yields

$$\mu = \sqrt{\frac{\log(n) - 2\log(q)}{n-1}} \pm \sqrt{\frac{2\log(q)(n-2) + n\log(n)}{n(n-1)}}. \qquad (17)$$

The left-hand term is the sample mean from equation (16), so the pair of $\mu$ in this equation always straddles the sample mean in equation (16). It also is clear that the right-hand term is larger than the left-hand term in equation (17), so the pair of $\mu$ also always straddles 0, our null-hypothesis value. Their difference is twice the right-hand term, and as $n$ goes to infinity the right-hand term goes to 0. Thus, the pair of $\mu$ may be made arbitrarily close to one another and therefore arbitrarily close to the null that they straddle. That is, we may conclude that for any criterial evidence-ratio level *q*, it is possible for a hypothetical $\mu_0$ to be <u>supported</u> at that level with another $\mu$ arbitrarily close to it that is <u>rejected</u> at the same level.

Is a better approach available? The nub of the difficulties revealed in this paper is the use of a point-valued null. If we must make a decision for or against the null, we could a priori select a small range around the null, as in Kruschke's recommendation[12] (Kruschke, 2011, 2018) a "region of practical equivalence" (ROPE). However, here we put it to a somewhat different use from his. We may use evidence-ratios $ER(\theta, x)$ to determine support and rejection regions and ascertain whether these lie within the ROPE. Considerations regarding the width and location of the ROPE are reviewed in Kruschke (2018).

If a support region containing the null hypothesis $\theta_0$ also is entirely contained in a ROPE around the null hypothesis value $\theta_0$, then we may conclude in favor of the null. In our Gaussian example, $ER(\mu, x) \geq 3$ when $0 \leq \mu \leq 0.07111$. If prior to obtaining our data we had decided that $-0.1 \leq \mu \leq 0.1$ constitutes our ROPE and our threshold is $q = 3$, then the resulting support region includes 0 and is contained by the interval [-0.1, 0.1], and we therefore could decide in favor of the null. If we want to be more cautious, we could also insist on additional conditions. For instance, we could require that the average $ER(\mu, x)$ within the ROPE exceed some threshold. Conversely, if one of the rejection regions entirely contains the ROPE then we can reject the null.

Conclusions

It seems wise to take the problems raised here seriously. They demonstrate that, depending on the nature of the alternative being compared to a point-valued null, it is possible for the same data to simultaneously constitute evidence for and against the null. To review the main findings in this paper:

1. For any BF threshold $q$ and sample mean $\bar{X}$, there is a value $n$ such that sample sizes larger than $n$ guarantee the occurrence of Lindley cases: Although the BF comparing H$_0$ against a conventional (vague) alternative exceeds $q$, nevertheless for some range of hypothetical $\mu$, a BF comparing H$_0$ against $\mu$ in that range falls below $1/q$.

2. For any such $\bar{X}$ and $q$, as $n$ increases the BF < $1/q$ range for $\mu$ will expand and the conventional BF also will increase. The evidence for and against the null will become stronger.

3. Although the threshold $\bar{X}$ moves closer to 0 with increasing $n$, the upper bound on the conventional BF for which Lindley cases occur nevertheless increases.

4. For any value of $q$ regarded by many researchers as "substantial", the probability of Lindley cases is non-negligible even when H$_0$ is true, and increases with sample size.

5. In a support interval setting, for any level of evidence-ratio, with increasing $n$ it is possible for a point-null $\mu = 0$ to be <u>supported</u> at that level with another $\mu$ arbitrarily close to it that is <u>rejected</u> at the same level.

6. Bayes factors comparing a point null against point alternatives are proportional to evidence-ratios for the point alternatives. Therefore, any criterion for determining a support or rejection region will correspond to a decisional criterion for a Bayes factor comparing a point null with a point alternative, and vice-versa.

The sample sizes used here for demonstrations are not outlandish, especially in this age of Big Data. Researchers working with large data-sets are likely to encounter Lindley cases if they find small effects. The results presented here suggest that, on its own, a Bayes factor comparing a point-valued null to a vague alternative may be of little utility in assessing the degree of support for a pointwise null hypothesis, and its utility declines with increasing sample size. They also indicate that using support and rejection regions for deciding whether to reject or confirm a point null can be problematic. Here, the nature of the prior will have an effect. Introducing the use of a ROPE into this decision making process shows some promise, at least for avoiding Lindley cases and some other pitfalls of point-null testing.

These suggestions are not the only ways that the problems raised here can be resolved, and it is beyond the scope of this paper to mount an exhaustive survey of potential resolutions. Instead, the main point of this paper has been the point out that there is an additional Achilles' heel in the Bayesian framework besides the choice of priors. Bayesian hypothesis testing also is vulnerable to attempting to "prove" a point-valued null, and that vulnerability becomes more apparent with larger samples.